\documentstyle[12pt]{article}
\def\p{\partial}
\def\tr{\,{\rm tr}\,}

\def\aor{\alpha_{1}^{\rm ort}}
\def\am{A_{\mu}}
\def\amt{A_{\mu}^{T}}
\def\an{A_{\nu}}
\def\ant{A_{\nu}^{T}}
\def\al{A_{\lambda}}

\def\alt{A_{\lambda}^{T}}
\def\ast{A_{\sigma}^{T}}
\def\ls{i\bar{\psi}\gamma^\mu(\partial_\mu +\am)\psi}

\def\be{\begin{equation}}
\def\ee{\end{equation}}
\def\bea{\begin{eqnarray}}
\def\eea{\end{eqnarray}}
\def\wz{Wess-Zumino }
\def\e{\epsilon}
\def\a{\alpha}
\title{\hfill SMI-33-94 \vspace{1.5cm}
{\bf \mbox{} \\ Canonical quantization of the degenerate WZ action including
chiral     interaction with gauge fields.
}}
\vspace{.7cm}
\author{ \mbox{}
\\ S.A.Frolov \thanks{Alexander von Humboldt fellow}
\mbox{} \\ \vspace{0.4cm} Section Physik, Munich University
\vspace{-0.5cm} \mbox{} \\ Theresienstr.37, 80333 Munich, Germany
\thanks{Permanent address:\ Steklov Mathematical Institute, Moscow}
\mbox{} \\ \\ \vspace{.6cm} A.A.Slavnov and C.Sochichiu \thanks{and Moscow
State University}
\vspace{-0.5cm} \mbox{} \\ Steklov Mathematical Institute
\vspace{-0.1cm} \mbox{}  \\ Vavilov st.42, GSP-1, 117966 Moscow, RUSSIA
\date{}}

\begin{document}

\maketitle
\vspace{4.5cm}
\begin{abstract}
Canonical quantization of the Wess-Zumino (WZ) model
including chiral interaction with gauge field is considered for the
case of a degenerate action. The two-dimensional SU(2) Yang-Mills model
and the four-dimensional SU(3) chiral gauge model proposed in the paper
\cite{fss} are studied in details. Gauge invariance of the quantum theory
is established at the formal level.

\end{abstract}

\newpage
\section {Introduction}

It is known that chiral gauge models suffer from anomalies \cite{Ad,Bar,GJ}.
The calculation of  the anomalies was performed both in the framework of
perturbation
 theory  and by algebraic and geometric methods \cite{St,Zu,Fa,FS84}. From
 algebraic point of view the anomaly corresponds to an infinitesimal 1-cocycle
 on a group $G$. The global 1-cocycle as was indicated by Faddeev and
Shatashvili \cite{Fa,FS84} is just the Wess--Zumino action
 \cite{WZ}, depending  on the chiral fields with values  in the group $G$.
The anomaly leads also to the appearence of an additional term in the
constraints commutator which is the infinitesimal 2-cocycle on the group
$G$ \cite{Fa,FS84,FS86,mic}. It was argued that it may change the physical
content
of the theory.

The particular form of anomaly, the corresponding Wess--Zumino term and
2-cocycle depend on the regularization used. Although the difference is a
local term it  may lead  to important physical consequences. In the two
dimensional case it was shown by Jackiw and  Rajaraman \cite{JR} that
different counterterms result in the different spectrum  of  the  model.

Recently we proposed a regularization of the four-dimensional chiral
$SU(N)$ Yang-Mills model, which preserves the gauge invariance with respect to
the $SO(N)$ subgroup, and have calculated the corresponding \wz action and the
infinitesimal 2-cocycle \cite{fss}. This "orthogonal" \wz action depends on the
chiral fields with values in the coset space $SU(N)/SO(N)$. For the $SU(3)$
model there are five chiral fields and, therefore, the corresponding
symplectic form is unavoidably degenerate. In the case of the "standard" \wz
action
the corresponding symplectic form is degenerate for the $SU(2k)$ models.
The degeneracy of the form
leads to the necessity of some modification in the quantization scheme of
anomalous models proposed by Faddeev and Shatashvili. In their approach
one should add to the original classical action the \wz term to restore
the quantum gauge invariance of the model and then to quantize the
modified action. The Hamiltonian quantization of the model results in the
appearence of the determinant of the symplectic form, if this form is
nondegenerate, in the integration measure as in the case of the "standard"
\wz action for the $SU(3)$ model. In the case of a degenerate symplectic
form one faces the problem of more careful Hamiltonian analysis of the
model.

In this paper we consider the canonical quantization of models
having a degenerate symplectic form on examples of
the two-dimensional chiral $SU(2)$ Yang-Mills model and the four-dimensional
chiral
$SU(3)$ gauge theory including the orthogonal \wz action.

In the two-dimensional case there is a family
of \wz actions, parametrized by one parameter $a$, and the choice of $a=1$
corresponds to the topological \wz action with the degenerate symplectic form.
As was mensioned by Shatashvili \cite{sh2d} this case differs from the others
and
requires a special analysis. We are going to use the standard canonical
approach
to the quantization of the model and therefore, we firstly need to rewrite the
\wz action which as is known includes a three-dimensional term as a
purely two-dimensional one. One can do it by using a special
parametrization of the $SU(2)$ group closely related to the
parametrization by the Euler angles. Of course two different
parametrizations lead to different \wz actions, but the difference is
equal to $2\pi n$, where $n$ is some integer, and does not influence  the
dynamics.  Then taking into account that the model has the gauge
invariance due to the presence of the \wz action we impose the light-cone
gauge and perform the canonical quantization of the theory. The light-cone
gauge is selected from other gauges because this gauge is
Lorentz-invariant and fermions do not interact with the vector fields
(more exactly, the effect of the interaction of fermions and vector fields
is expressed by the \wz action). We show that the model has three primary
constraints and one secondary constraint. This secondary constraint and
the determinant of the matrix of the Poisson brackets of the constraints
are gauge-invariant and,
therefore, the integration measure in the path integral is gauge-invariant
too.
Using the gauge invariance of the measure one can show in the usual way
\cite{FS86}
that the Gauss-law constraints
form the $SU(2)$ gauge algebra and that one can select the physical sector
imposing these constraints on the state vectors. This justifies the possibility
of
imposing the gauge condition before the quantization of the model.
Solving the primary and
secondary constraints one can show that the theory describes three vector
fields,
one boson field and one free fermion. Due to the Gauss-law constraints only the
boson field and the fermion are physical.

In the third section we discuss  in the same manner the four-dimensional
model with the orthogonal \wz action. We  indicate  a parametrization of the
coset
space $SU(N)/SO(N)$ reducing the \wz action to a pure four-dimensional one.
A simple modification of this parametrization can be also used for the standard
\wz action. Then we calculate the five primary constraints and the symplectic
form of
the model and show that there is only one null vector of the form and that the
corresponding secondary constraint is gauge-invariant up to some factor. The
primary
and secondary
constraints form a set of second-class constraints and
therefore adding the
orthogonal \wz action to the gauge theory one gets a system with two new
physical
degrees of freedom. In the case of the standard \wz action one gets four new
physical degrees of freedom and thus these models  differ crucially  from
each other in spite of the fact that the difference of the two \wz actions
is a local trivial 1-cocycle.

\section{Two-dimensional model}

We consider the chiral $SU(2)$ Yang-Mills theory described by the Lagrangian
\be
S_{YM}=\int d^2x\,(-\frac {1}{4e^{2}} (F_{\mu\nu}^{a})^{2} +
i\psi_{L}^{+}(\partial_{+}+A_{+})\psi_{L}),
\label{1}
\ee
where $\partial_{+}=\partial_{0}+\partial_{1}; \qquad A_{+}=A_{0}+A_{1}$.

On the classical level this action possesses the usual gauge invariance,
however as is well-known quantum corrections violate this invariance. To
restore the gauge invariance one can following Faddeev and Shatashvili
\cite{FS86}
add to the action (1) the corresponding \wz action, which in our case looks as
follows:
\be
S_{WZ}=-\frac {1}{12\pi}\int_{M^{+}} d^3x\,\e^{ijk}\tr g_{i}g_{j}g_{k}+
\frac {1}{4\pi}\int d^2x\,\e^{\mu\nu}\tr g_{\mu}A_{\nu}
\label{2}
\ee
Here $g_{i}=\partial_{i}gg^{-1},\quad g\in SU(2),\quad\e^{ijk}$ and
$\e^{\mu\nu}$ are
antisymmetric tensor fields, $M^{+}$ is a three-dimensional manifold whose
boundary is the usual two-dimensional space.

Then one should find a gauge-invariant measure in the path integral over all
fields including the chiral fields $g$. To define the measure we impose the
light-cone gauge $A_{+}=0$ and use the standard canonical formalism and, then,
check that the measure obtained is really gauge-invariant, justifying thus
the possibility of imposing the gauge condition before the quantization.

The only problem in applying the canonical quantization is the
three-dimensional
term in the \wz action (2). However it is known that this term depends only on
the values of the chiral field $g$ on the two-dimensional boundary (more
exactly
by $mod \, 2\pi$) and therefore one  can choose  such a parametrization of
the field $g$ in which this term can be written exlicitely as a
two-dimensional one.  We use the parametrization of the $SU(2)$ group by
the fields $\phi^{A}$, satisfying the following condition:
\be
\tr
(g_{A}g_{B}g_{C})=6\pi\e_{ABC}, \label{3}
\ee
where $g_{A}=\frac {\partial
g}{\partial \phi^{A}}g^{-1}=\partial_{a}gg^{-1}$ is a right-invariant
vector field on the $SU(2)$ group.

In terms of the fields $\phi^{A}$ any right-invariant current $g_{i}$ can be
expressed by the following formula:
\be
g_{i}=\frac {\partial g}{\partial x^{i}}g^{-1}=g_{A}\partial_{i}\phi^{A}
\label{4}
\ee
Due to the condition (3) the Haar measure $dgg^{-1}$ on the $SU(2)$ group is
proportional to $d\phi^{1}d\phi^{2}d\phi^{3}$.

Using the parametrization by the fields $\phi^{A}$ and imposing the light-cone
gauge one can rewrite the sum of (1) and (2) as follows:
\be
S=\int d^2x\,(\frac {1}{2e^{2}} (\partial_{+}A^{a})^{2} +
\frac {1}{2}\e_{ABC}\e^{\mu\nu}\phi^{A}\partial_{\mu}\phi^{B}\partial_{\nu}
\phi^{C}
+\frac {1}{4\pi}\tr (g_{A}A)\partial_{+}\phi^{A})
\label{5}
\ee
Here $A=\frac 12 (A_{1}-A_{0})$ and we have omitted the fermion part of the
action
because fermions do not interact with the vector fields in the light-cone
gauge.

Introducing the canonically-conjugated momenta for the fields $A$ and
$\phi^{A}$
one can present the action (5) in an equivalent form:
\bea
S&=&\int d^2x\,(E_{a}\partial_{0}A^{a}+p_{A}\partial_{0}\phi^{A}+
E_{a}\partial_{1}A^{a}-\frac 12 (E_{a})^{2} +
\frac {1}{4\pi}\tr (g_{A}A)\partial_{1}\phi^{A}\nonumber\\
&+&\lambda^{A}(p_{A}+\e_{ABC}\phi^{B}\partial_{1}\phi^{C}-
\frac {1}{4\pi}\tr (g_{A}A))
\label{6}
\eea
{}From (6) one can conclude that
\be
H=-E_{a}\partial_{1}A^{a}+\frac 12 (E_{a})^{2}-
\frac {1}{4\pi}\tr (g_{A}A)\partial_{1}\phi^{A}
\label{7}
\ee
is the Hamiltonian and
\be
C_{A}=p_{A}+\e_{ABC}\phi^{B}\partial_{1}\phi^{C}-\frac {1}{4\pi}\tr (g_{A}A)
\label{8}
\ee
are the primary constraints of the model.

The next step in the canonical quantization is the calculation of  secondary
constraints. The simplest way seems to be to find all null-vectors of the
matrix of the Poisson brackets of the primary constraints. Then for every
null-vector
$e_{\alpha}^{a}$ one can form a linear combination of the primary constraints
$C_{\alpha}=C_{A}e_{\alpha}^{A}$, which commutes with all primary constraints
on the constraints surface. The secondary constraints are then given by the
the Poisson brackets of $C_{\alpha}$ and the Hamiltonian $H$.

In our case the matrix of the Poisson brackets the primary constraints is
equal to:
\bea
\Omega_{AB} (x^{1},y^{1})&=&\{ C_{A}(x^{1}),C_{B}(y^{1})\} =
\Omega_{AB} (x^{1})\delta(x^{1}-y^{1})\nonumber\\ &=&
\frac {1}{4\pi}\tr ([g_{A},g_{B}](g_{1}(x^{1})+A(x^{1})))\delta(x^{1}-y^{1})
\label{9}
\eea
This matrix is ultralocal and in fact coincides with the symplectic form for
the
\wz action. There is only one null-vector of $\Omega_{AB} $ (in every space
point)
equal to
\be
e^{A}(x^{1})=\frac {1}{4\pi}\e^{ABC}\tr ([g_{B},g_{C}](g_{1}(x^{1})+A(x^{1}))=
\e^{ABC}\Omega_{BC} (x^{1})
\label{10}
\ee
Calculating the Poisson bracket of the constraint
 $\widetilde{C}(x^{1})= C_{A}(x^{1})e^{A}(x^{1})$
and $H$ one gets up to the primary constraints the secondary constraint:
\be
C(x^{1})=4\pi\{ H, \widetilde{C}(x^{1})\} =\tr (E(x^{1})(g_{1}(x^{1})+A(x^{1}))
\label{11}
\ee
 In this equation we omited the term proportional to $C_A$.
The primary constraints $C_{A}(x^{1})$ and the secondary constraint $C(x^{1})$
form a set of second-class constraints and the matrix of the Poisson brackets
of
the constraints is equal to:
\be
M(x^{1},y^{1})=\left( \begin{array}{cc} \Omega_{AB} (x^{1},y^{1}) &
v_{A}(x^{1},y^{1}) \\
-v_{B}(y^{1},x^{1}) & 0 \end{array} \right)
\label{12}
\ee
where
\bea
v_{A}(x^{1},y^{1})&=&\{ C_{A}(x^{1}),C(y^{1})\} \nonumber\\ &=&
\tr (g_{A}(\partial_{1} E-[g_{1},E] - \frac {1}{4\pi} (g_{1}+A)))
\delta(x^{1}-y^{1})\nonumber\\ &\quad &-\tr (g_{A}E(x^{1}))\partial_{1}^{y}
\delta(x^{1}-y^{1})
\label{13}
\eea
It is not difficult to show that the determinant of the matrix $M$ is equal
to
\be
\det M=(\det \e^{ABC}\Omega_{AB}v_{C})^{2}=(\det e^{A}v_{A})^{2}
\label{14}
\ee
and
\be
e^{A}v_{A}(x^{1},y^{1})=
\tr ((g_{1}+A)(\nabla_{1}E-\frac {1}{4\pi}
(g_{1}+A)))\delta(x^{1}-y^{1})
\label{15}
\ee
up to the secondary constraint $C(x^{1})$.
Now one can write the expression for the generating functional of the
model:
\bea
Z&=&\int \,DADED\phi DpD\psi (\det M)^{\frac 12}\delta(C)\delta(C_{A})
\nonumber\\
&\quad &\hbox{exp}\{ i\int d^2x\,(E_{a}\partial_{+}A^{a}+p_{A}\partial_{+}
\phi^{A}
-\frac 12 (E_{a})^{2} + i\psi_{L}^{+}\partial_{+}\psi_{L})\}
\label{16}
\eea
Integrating over $p_{A}$ and introducing the integration over $A_{+}$ in
the path integral one gets
\be
Z=\int \,DA_{\mu}DED\phi D\psi (\det M)^{\frac 12}\delta(C)\delta(A_{+})
\hbox{exp}\{ i(S_{YM}+S_{WZ}) \}
\label{17}
\ee
and we use in eq.(17) the following expressions for $e^{A}v_{A}$ and $C$
due to the fact that $A=A_{1}$ if $A_{+}=0$
\be
e^{A}v_{A}(x^{1},y^{1})=\tr ((g_{1}+A_{1})(\nabla_{1}E-\frac {1}{4\pi}
(g_{1}+A_{1})))\delta(x^{1}-y^{1})
\label{18}
\ee

\be
C(x^{1})=\tr (E(x^{1})(g_{1}(x^{1})+A_{1}(x^{1}))
\label{19}
\ee
It is obvious from eqs.(18,19) that the integration measure in eq.(17) is
gauge-invariant  apart from  the gauge-fixing condition and the fermion
measure and therefore one can easily show that the modified Gauss-law
constraints form the $SU(2)$ gauge algebra:  \be
 [G_{a}(x^{1}),G_{b}(y^{1})]=i\e_{abc} G_{c}(x^{1}) \delta(x^{1}-y^{1})
\label{20}
\ee
where
\be
G(x^{1})=\nabla_{1}E(x^{1})-\frac {1}{4\pi}g_{1}(x^{1})+j_{0}(x^{1})
\label{21}
\ee
Due to eq.(20) one can select the physical subspace imposing the condition
$G_{a}|\Psi >=0$ on the state vectors.
The number of the physical degrees of freedom can be now easily calculated.
All vector fields are unphysical due to the Gauss-law constraints and there
is only one physical degree of freedom for three chiral fields $\phi_{a}$
due to the four second-class constraints.

\section{Four-dimensional model}

In this section we present similar results for the chiral $SU(3)$
Yang-Mills theory. The
complete action of the model is described by the sum of the Yang-Mills
action and the orthogonal \wz action (we use notations
of \cite{fss}):
\be
S=\int d^4x\,(-\frac {1}{4e^{2}} (F_{\mu\nu}^{a})^{2}
+ \ls )+\aor (A,s) \label{22} \ee \bea \aor&=&\int d^4x\,[\frac
12d^{-1}\kappa (s)-\frac {i}{48\pi^2} \epsilon^{\mu \nu \lambda \sigma
}\tr[(\partial_\mu \an \al + \am \partial_\nu \al +\am \an \al-\nonumber
\\ & &-\frac 12 \partial_\mu \an s\alt s^{-1}- \frac 12 s \amt s^{-1}
\partial_\nu \al- \am s\ant s^{-1}\al)s_\sigma-\nonumber \\ & &-\frac 12
\am s_\nu \al s_\sigma +\frac 12 (s\amt s^{-1}\an-\am s\ant
s^{-1})s_\lambda s_\sigma - \am s_\nu s_\lambda s_\sigma \nonumber \\
& &+\partial_\mu \an \al s\ast s^{-1}+ \am \partial_\nu \al s\ast s^{-1} +
\am \an \al s\ast s^{-1}-\nonumber \\
& &-\frac 14 \am s \ant s^{-1}\al s\ast s^{-1}-\alpha_0 (A)]]
\label{23}
\eea
where $\psi \equiv \frac{1}{2}(1+\gamma_{5})\psi$
is a chiral  fermion in the fundamental representation, $s$ is a symmetric
unitary matrix, parametrizing the coset space $SU(N)/SO(N)$
$s_\mu=\partial_\mu ss^{-1}$
\bea
\alpha_0 (A)&=&-\frac {i}{48\pi^2}\int  d^4x\, \epsilon^{\mu \nu
\lambda \sigma } \tr( \am \an \al \ast-\frac 14 \am \ant  \al \ast
+\nonumber\\
& & + \partial_\mu \an \al \ast + \am \partial_\nu \al \ast )
\nonumber
\eea
and
\bea
\int d^4xd^{-1}\kappa (s) \equiv
-\frac{i}{240\pi^2}\int_{M_5} d^5x \,\epsilon^{pqrst} \tr{(s_p s_q s_r s_s
s_t)}
\nonumber
\eea
In the last equation the  integration goes  over a
five-dimensional manifold whose boundary is the usual  four-dimensional
space.

The action (22) is gauge-invariant on the quantum level and the gauge
group transforms the coordinates $s$ in the following manner:
\be
s\rightarrow  g^{-1}sg^{-1,T}.
\label{24}
\ee
The condition, that the \wz action $\aor (A,s)$ is a 1-cocycle can be
written in our case as follows:
\be
\aor (A^{h},h^{-1}sh^{-1,T})=\aor (A,s)-\aor (A,hh^{T})\qquad (mod\, 2\pi)
\label{25}
\ee
One can easily see that $\aor (A,s)$ is gauge-invariant with respect to the
$SO(N)$ subgroup of the $SU(N)$ group.

To apply the canonical formalism to the model one needs, as was mentioned
in the Introduction, to reduce the five-dimensional term of the \wz action
to a four-dimensional one. To do it one can use the fact that any symmetric
unitary matrix can be represented in the following form:
\be
s=\omega D\omega^{T}
\label{26}
\ee
where $\omega$ is an orthogonal matrix $\omega\omega^{T}=1$ and $D$ is a
diagonal unitary matrix.

Using this representation and eq.(25) one can show the validity of the
following equation:
\be
\aor (A,\omega D\omega^{T})=\aor (A^{\omega}, D)
\label{27}
\ee
The five-dimensional term is equal to zero for any diagonal matrix and
therefore the parametrization (26) solves the problem of reducing the
\wz action to a four-dimensional form.

Let us now represent the matrix $D$ in the form $D=\hbox{e} ^{u^{\a}T_{\a}}$,
where
matrices $T_{\a}$ belong to the Cartan subalgebra of the  $su(N)$
algebra, and use an arbitrary parametrization of the $SO(N)$ group by
fields $\phi^{A}$. Then introducing the canonically-conjugated momenta for
the fields $A_{i}$, $\phi^{A}$ and $u^{\a}$ and imposing the temporal
gauge $A_{0}=0$ one can rewrite the action (22) as follows:
\bea
S&=&\int d^4 x (\Pi^i_a \p_0 A_i^a + p_A \p_0 \phi^A + \pi_\alpha
\partial_0 u^\alpha - \frac{1}{2} (\Pi^i_a-\Delta E^i_a)^2-
\nonumber \\
&& \frac{1}{4} (F_{ij}^a)^2 + \lambda^\alpha C_\alpha+ \lambda^A C_A +
{\cal L}_\psi )
%\nonumber \\ &&\nonumber \\ &&\nonumber \\ &&
\label{28}
\eea
where
\bea
\Delta E^i_a&=&-\frac{i}{48\pi^2}\e^{ijk} \tr (T_a \omega (\{A_j^\omega,
u_k \}-\frac{1}{2} \{DA_j^{\omega,T}D^{-1},u_k \}+
\nonumber \\
&& + \{A_j^\omega, DA_k^{\omega, T} D^{-1} \} - \{A_j^\omega, A_k^{\omega,
T} \})\omega^{-1})
\nonumber \\
A_i^\omega&=& \omega^{-1}A_i \omega + \omega^{-1} \p \omega;\quad u_i=\p_i
u=\p_i DD^{-1}
%\nonumber \\ &&\nonumber \\ &&
\label{29}
\eea
and
$C_{p}=(C_{\a},C_{a})$ are the primary constraints of the model
\bea
C_\alpha&=&\pi_\alpha - \frac{i}{48 \pi^2} \e^{ijk} \tr T_\alpha (\{\p_i
A_j^\omega ,A_k^\omega \} + A_i^\omega A_j^\omega A_k^\omega -
\nonumber \\
&&-\frac{1}{2} \{\p_i A_j^\omega, DA_k^{\omega, T} D^{-1} \}- A_i^\omega
DA_i^{\omega, T} D^{-1} A_k^\omega -A_i^\omega u_j A_k^\omega
%\nonumber \\ &&\nonumber \\ &&
\label{30}
\eea
\bea
C_A&=&p_A+ \frac{i}{48 \pi^2} \e^{ijk} \tr \omega_A ( \{\p_i A_j^\omega,
u_k \} - \frac{1}{2} D\{\p_i A_j^{\omega, T} , u_k \} D^{-1}+
\nonumber \\
&+&A_i^\omega u_j A_k^\omega -D^{-1} A_i^\omega u_j A_k^\omega D -
DA_i^{\omega, T}D^{-1} A_j^\omega u_k -u_k A_i^\omega DA_j^{\omega ,T}
D^{-1}+
\nonumber \\
&+& \frac{1}{2} [A_i^\omega, \{DA_j^{\omega, T} D^{-1} , u_k \} ] - u_i
A_j^\omega u_k+ \{ \p_i A_j^\omega , DA_k^{\omega , T}D^{-1} - A_k^{\omega
, T} \} +
\nonumber \\
&+& D^{-1} \{A_k^\omega , \p_i A_j^\omega \} D - \{A_k^\omega , \p_i
A_j^\omega \} +D^{-1} A_i^\omega A_j^\omega A_k^\omega D -  A_i^\omega
A_j^\omega A_k^\omega
\nonumber \\
&-& D^{-1}A_i^\omega DA_j^{\omega, T}D^{-1} A_k^\omega D + A_i^\omega
DA_j^{\omega ,T}D^{-1} A_k^\omega)
%\nonumber \\ &&
\label{31}
\eea
As was mentioned
above, the matrix of the Poisson brackets of the primary constraints
coincides with the symplectic form and is equal to:
\bea
\Omega_{pq}({\bf x,y })&=& \{C_p({\bf x}),C_q({\bf y}) \} =
\Omega_{pq}({\bf x}) \delta({\bf x-y})
\nonumber \\
\Omega_{pq}({\bf x})&=& \frac{i}{96 \pi^2} \e^{ijk} \tr ([s_p , s_q](
\frac{1}{2} \{\widetilde{A}_i , \widetilde{F}_{jk} \}- \widetilde{A}_i
\widetilde{A}_k
\widetilde{A}_k )+
\nonumber \\
&+&s_p( \frac{1}{2} \widetilde{F}_{ij} -  \widetilde{A}_i
\widetilde{A}_j)s_q \widetilde{A}_k
s_q( \frac{1}{2} \widetilde{F}_{ij} -  \widetilde{A}_i
\widetilde{A}_j)s_p \widetilde{A}_k)
%\nonumber \\ &&\nonumber \\ &&
\label{32}
\eea
where
\bea
s_\alpha&=&\frac{\p s}{\p u^\alpha}s^{-1}=\omega \lambda_\alpha
\omega^{-1}; \quad s_A=\frac{\p s}{\p \phi^A}s^{-1}=\omega (
\omega_A-D\omega_A D^{-1})\omega^{-1};
\nonumber \\
&&\omega_A=\omega^{-1}\frac{\p \omega}{\p \phi^A}
\label{33}
\eea
and
\bea
\widetilde{F}_{ij}&=& F_{ij}-sF_{ij}^T s^{-1}; \quad \widetilde{A}_i=
A_i+sA_i^T s^{-1}+ s_i
%\nonumber \\ &&
\label{34}
\eea
For the $SU(3)$ theory
the only null vector of the symplectic form is equal to:
\bea
e^p({\bf x})&=&\e^{pqrst} \Omega_{qr}({\bf x}) \Omega_{st} ({\bf y})
%\nonumber \\ &&
\label{35}
\eea
As before the secondary constraint is
given by the Poisson bracket  of the constraint
$\widetilde{C}({\bf x})=C_{p}({\bf
x})e^{p}({\bf x})$ and the Hamiltonian $H=\int d^3 x (\frac{1}{2}(\Pi_a^i-
\Delta E_a^i)^2 + \frac{1}{4} (F_{ij}^a)^2)$:
\bea
C({\bf x})&=& \{C_p({\bf x}) e^p({\bf x}), H \} \sim \e^{pqrst}
R_p \Omega_{qr}({\bf x}) \Omega_{st} ({\bf y})
%\nonumber \\ &&
\label{36}
\eea
where
\bea
R_p&=&\e^{ijk} \tr s_p ( \{E_i , F_{jk}- \frac{1}{2} sF_{jk}^T
s^{-1}-\widetilde{A}_j \widetilde{A}_k \} + \widetilde{A}_i E_j
\widetilde{A}_k)
%\nonumber \\ &&\nonumber \\ &&
\label{37}
\eea
It is not
difficult to show that the secondary constraint transforms under the gauge
transformation as follows:
\be
C({\bf x}) \rightarrow  \det \left( \frac{ \p
\phi^p}{\p \widetilde{\phi^q}} \right) C({\bf x})
\ee
where $\phi^{p}$ are the coordinates of the point $s$
on the coset space $SU(3)/SO(3)$ ($u^{\a}$ and $\phi^{A}$ in our case) and
$\tilde\phi ^{p}$ are the coordinates of the gauge-transformed point
$g^{-1}sg^{-1,T}$. In other words the function $\tilde\phi(\phi)$ defines the
change of the field $\phi^{p}$ under to the gauge transformation.  To prove
eq.(38) one should take into account the following transformation law of
$s_{p}=\frac {\partial s} {\partial\phi^{p}} s^{-1}$:
\bea
s_p&\rightarrow &
g^{-1} s_q g \frac{\p \phi^q}{\p \widetilde{\phi^p}}
%\nonumber \\ &&
\label{39}
\eea
The five
primary constraints $C_{p}({\bf x})$ and the secondary constraint $C({\bf
x})$ form a set of second-class constraints with the following matrix of
the Poisson brackets of the constraints:  \be M({\bf x},{\bf y})=\left(
\begin{array}{cc} \Omega_{pq} ({\bf x},{\bf y}) & v_{p}({\bf x},{\bf y})
\\ -v_{q}({\bf y},{\bf x}) & v({\bf x},{\bf y}) \end{array} \right)
\label{40}
\ee
where
\be
v_{p}({\bf x},{\bf y})=\{
C_{p}({\bf x}),C({\bf y})\} ,\qquad v({\bf x},{\bf y})=\{ C({\bf
x}),C({\bf y})\}
\label{41}
\ee
The explicit formulas for $v_{p}$ and $v$
are rather complicated, but one can show that the matrix $M({\bf x},{\bf
y})$ has the following law of the gauge transformation:
\bea
M({\bf x},{\bf y})
&\rightarrow & \left( \begin{array}{cc} \frac{\p \phi^p}{\p
\widetilde{\phi}^r}({\bf x}) & 0 \\
0 & det \left( \frac{\p \phi}{\p \widetilde{\phi}}({\bf x})\right)
\end{array} \right) \left( \begin{array}{cc} \Omega_{pq} ({\bf x},{\bf y})
& v_{p}({\bf x},{\bf y}) \\ -v_{q}({\bf y},{\bf x}) & v({\bf x},{\bf y})
\end{array} \right) \times
\nonumber \\
&& \quad
\left( \begin{array}{cc} \frac{\p \phi^q}{\p
\widetilde{\phi}^s}({\bf y}) & 0 \\
0 & det \left( \frac{\p \phi}{\p \widetilde{\phi}}({\bf y}) \right)
\end{array} \right)
%\nonumber \\ &&\nonumber \\ &&
\label{42}
\eea
Due to
eq.(42) $(\det M)^{\frac 12}$ transforms as follows:  \bea
(detM)^{\frac{1}{2}}& \rightarrow & \left(det \frac{\p \phi}{\p
\widetilde{\phi}}\right)^{2} (detM)^{\frac{1}{2}}
\nonumber \\ && \label{43} \eea Now we can prove
the gauge invariance of the integration measure in the path integral for
the generating functional:
\be Z=\int \,DA_{\mu}DED\phi D\psi (\det
M)^{\frac 12}\delta(C)\delta(A_{0}) \hbox{exp}\{ i(S_{YM}+S_{WZ}) \}
\label{44}
\ee
Taking into account eqs.(38) and (43) one can easily see
the gauge invariance of the measure $D\phi (\det M)^{\frac 12}\delta(C)$
and therefore the invariance of the integration measure in eq.(44) apart
from the gauge-fixing condition and the fermion measure. Thus we have
proved the possibility of imposing the gauge condition before the
quantization and of selecting the physical subspace by the Gauss-law
constraints.

The number of the physical degrees of freedom can be now easily
calculated. Due to the Gauss-law constraints there are $2\times 8$ vector
degrees of freedom (8 is the dimension of $SU(3)$) and due to the six
second-class constraints there are two boson degrees of freedom. Let us
remind that in the case of the standard \wz action one would get four
boson degrees of freedom and thus these models differ crucially from each
other in spite of the fact that the difference between these \wz actions
is a local trivial 1-cocycle.  Let us finally note that one could use such
a parametrization of the coset space $SU(3)/SO(3)$ that the invariant
measure on the space is proportional to $d\phi^{1}...d\phi^{5}$. In this
case the secondary constraint and $\det M$ are gauge-invariant.

{\bf Discussion}
In this paper we proved that the formal proof of the gauge invariance of
the \wz model including chiral interaction with gauge fields remains valid
in the case of degenerate symplectic form as well. By the usual arguments
one can prove that the Gauss law commutator has a standard form and
therefore gauge invariance is restored at the quantum level.

{\bf Acknowledgements:} One of the authors (S.F.) would like to thank
Professor J.Wess for kind hospitality and the Alexander von Humboldt
Foundation for the support. This work has been supported in part by
ISF-grant MNB000 and by the Russian Fund for Fundamental Studies under
grant number 94-01-00300a.

\end{document}